**Towards Increased Reliability, Transparency and Accessibility in Crosslinking Mass Spectrometry**

Alexander Leitner[1,2,*], Alexandre M.J.J. Bonvin [X], Christoph H. Borchers, Robert J. Chalkley, Julia Chamot-Rooke, Colin W. Combe, Jürgen Cox, Meng-Qiu Dong, Lutz Fischer, Michael Götze, Fabio C. Gozzo, Albert J. R. Heck, Michael R. Hoopmann, Lan Huang, Yasushi Ishihama, Andrew R. Jones, Nir Kalisman, Oliver Kohlbacher, Karl Mechtler, Robert L. Moritz, Eugen Netz, Petr Novak, Evgeniy Petrotchenko, Andrej Sali, Richard A. Scheltema, Carla Schmidt, David Schriemer, Andrea Sinz, Frank Sobott, Florian Stengel, Konstantinos Thalassinos, Henning Urlaub, Rosa Viner, Juan A. Vizcaino, Marc R. Wilkins, Juri Rappsilber[3,4,*]

[1] Department of Biology, Institute of Molecular Systems Biology, ETH Zurich, 8093 Zurich, Switzerland

[2] Lead author

[3] Bioanalytics, Institute of Biotechnology, Technische Universität Berlin, 13355 Berlin, Germany

[4] Wellcome Centre for Cell Biology, School of Biological Sciences, University of Edinburgh, Edinburgh EH9 3BF, Scotland, United Kingdom

[X] A complete list of all other author affiliations, along with the author's ORCiD IDs, is provided in a separate spreadsheet

* Correspondence: leitner@imsb.biol.ethz.ch (A.L.), juri.rappsilber@tu-berlin.de (J.R.)

**Summary**

Crosslinking mass spectrometry (Crosslinking MS) has substantially matured as a method over the last two decades through parallel development in multiple labs, demonstrating its applicability for protein structure determination, conformation analysis and mapping protein interactions in complex mixtures. Crosslinking MS has become a much-appreciated and routinely applied tool especially in structural biology. Therefore, it is timely that the community commits to the development of methodological and reporting standards. This white paper builds on an open process comprising a number of events at community conferences since 2015 and identifies aspects of Crosslinking MS for which guidelines should be developed as part of a Crosslinking MS standards initiative.

**1. Introduction**

Crosslinking for structural analysis goes back at least as far as 1958, when the topology of insulin was investigated by help of a crosslinking reagent (Zahn and Meienhofer, 1958). Introducing MS for the detection of crosslinks (crosslinking mass spectrometry, here abbreviated as Crosslinking MS, but also known as XL-MS, CXMS, or CLMS) led to increased accuracy of identifying which pairs of proteins were linked together in heteromeric complexes and increased resolution by revealing the identity of the linked residues and thus the interaction regions within these proteins. Technical progress, including the wide variety of parallel developments, and biological applications have been reviewed extensively in recent years (Leitner et al., 2016; O'Reilly and Rappsilber, 2018; Sinz, 2018; Steigenberger et al., 2020; Yu and Huang, 2018).

Encouraged by the worldwide Protein Data Bank (wwPDB) Task Force for Integrative/Hybrid Methods (Berman et al., 2019; Sali et al., 2015) to provide experimentalists and modellers with a stable access point to crosslinking data, a first open gathering on standards in the Crosslinking MS field took place at 14[th] Human Proteome Organization World Congress - HUPO 2015 in Vancouver (Canada). This effort was carried forward into an open podium discussion at the 5[th] Symposium on Structural Proteomics in Halle/Saale (Germany) later that year. As a result, at the HUPO-Proteomics Standards Initiative (PSI) meeting 2016 in Ghent (Belgium), mzIdentML, the proteomics data standard for peptide/protein identification information, embraced crosslinking data from version 1.2 (Vizcaíno et al., 2017). Following a closed meeting of senior investigators at the 7[th] Symposium on Structural Proteomics in Vienna (Austria), 2017, again an open podium discussion took place at the 8[th] Symposium on Structural Proteomics in Berlin (Germany), 2018. A first community-wide, comparative Crosslinking MS study was published in 2019 (Iacobucci et al., 2019), organised within the European Union COST Action BM1403 as an initiative to develop activities in structural proteomics at large, which includes Crosslinking MS. Discussions were continued during three meetings in 2019: the American Society for Mass Spectrometry (ASMS) Sanibel Conference 2019 entitled "Chemical Cross-linking and Covalent Labeling: From Proteins to Cellular Networks", the Dagstuhl Seminar 19351 "Computational Proteomics", and the 18th Human Proteome Organization World Congress - HUPO 2019 in Adelaide. These efforts were brought together at the 9th Symposium on Structural Proteomics in Göttingen (Germany) later that year. A questionnaire on standardisation of Crosslinking MS was circulated among the participants of the meeting. Also, a discussion group formed with participation of 20 research labs and companies with strong interests in Crosslinking MS and thus strong representation of the field that formulated challenges and recommendations for the field. These were publicly discussed within the conference at the end of the meeting. The resulting draft document was then circulated to participants and additionally to labs in Crosslinking MS that were not represented in Göttingen, to provide them with an opportunity to participate. Following the discussions at these meetings, this white paper is now supported by about 30 academic laboratories and companies engaged in developing, applying and supporting Crosslinking MS and thus a substantial fraction of the field.

## 2. Crosslinking at the interface of proteomics and structural biology

> Definition of Crosslinking MS: Non-covalent interactions or proximities within or between biomolecules are covalently fixed for their detection in an otherwise dissociative analytical process involving a mass spectrometer.

Crosslinking MS shares some similarities to conventional structural biology techniques, but also has some distinct features. For example, structural dynamics in solution is not appropriately reflected in static structures obtained by X-ray crystallography. NMR spectroscopy and to some extent (cryo-)electron microscopy are able to reveal ensembles of conformational states. Crosslinking data also reflect such solution phase dynamics and are often able to provide crucial contact information about flexible regions in proteins that remain inaccessible to EM or crystallography and are therefore absent in many deposited structures and models. Because of this complementarity to established structural methods, Crosslinking

MS has gained acceptance in the structural biology community and any efforts towards standardisation should also align with best practices in that field.

Crosslinking MS is most intimately connected to a wide range of applications in structural biology, but is at the same time rooted in MS-based proteomics, two fields where substantial efforts for standardisation and harmonisation have developed in the last decades (Berman et al., 2006; Burley et al., 2017; Deutsch et al., 2017; Lawson et al., 2011; Montelione et al., 2013; Sali et al., 2015; Schwede et al., 2009; Trewhella et al., 2013, 2017; Vallat et al., 2018). However, the primary data output from Crosslinking MS experiments already differs substantially from conventional protein identification and quantification workflows in proteomics. Instead of identifying single peptide chains that then jointly identify proteins, the main types of identifications in Crosslinking MS are pairs of covalently connected, i.e. crosslinked peptides, whereby these peptides may originate from the same or two different protein(s). These are then combined into sets of crosslinking sites (pairs of crosslinked residues) or, for samples of higher complexity, pairs of interacting proteins. Crosslinking involves the use of one or more of a large variety of crosslinking reagents, data acquisition strategies and data analysis approaches. Inevitably, this diversity will pose particular challenges when it comes to standardising workflows and data formats, and for introducing reporting guidelines that the community adheres to.

Any initiative that attempts to establish standards and guidelines in relation to crosslinking needs to capture most of today's diverse crosslinking community. We are in the fortunate position to witness the growth of this community and one might broadly define members of the crosslinking community as researchers or labs that (a) develop crosslinking chemistries, workflows, software etc.; and those that (b) apply crosslinking to address questions in structural biology, molecular biology, systems biology and so on. In some cases, the focus of research groups in the Crosslinking MS field may of course cover both directions. In addition, crosslinking methods are not restricted to the study of protein-protein interactions and protein conformations, but may also include interactions with other classes of biomolecules, including other biopolymers and small molecules. In fact, also natural processes can lead to crosslinks and these products can be analysed by the tools of Crosslinking MS.

To increase transparency and access to results, the crosslinking community has already established some organisational liaisons, including most prominently with proteomics data repositories (ProteomeXchange Consortium (PXC) partners (Deutsch et al., 2020), in particular PRIDE (Perez-Riverol et al., 2019) and the jPOST (Moriya et al., 2019)), and repositories for integrative/hybrid structural biology (particularly PDB-Dev, a prototype repository of the wwPDB for integrative structures (Burley et al., 2017; Vallat et al., 2018)). At the moment, these two types of repositories cover different parts of the crosslinking workflow and are not yet interconnected. In the following, we will outline the different steps of this workflow and how researchers from different communities may benefit from increased reliability, transparency and access.

## 3. The crosslinking experiment: from sample to shared data

**FIGURE 1** outlines the different steps of a crosslinking experiment and highlights specific steps of the procedure. For the sake of this discussion, we will assume that independent of the sample type and the specific chemistry involved, a protein or protein mixture has been crosslinked, subsequently digested into peptides using one or more proteases and the resulting peptide mixture has been analysed by liquid chromatography coupled to tandem MS (LC-MS/MS). This experimental workflow can deliver (a) sites of crosslinks that may inform modelling of a protein or protein complex structure and (b) information on which proteins were linked and thus interacting in a possibly highly complex biological mixture. The raw/primary data emerging from an LC-MS/MS experiment are considered the starting point of the data analysis pipeline. Primary data that is generated in vendor specific formats may first be converted into open file formats suitable for database searching (although some programs may be able to work with vendor specific formats directly). Either way, a peak list is generated that corresponds to the experimentally acquired MS/MS spectra that are searched against a protein sequence database of interest. This database may contain anything from a few protein sequences of interest up to a whole proteome database.

Successful matches to the experimental spectra are designated either peptide-spectrum matches (PSMs) or crosslink-spectrum matches (CSM, XSM). These matches correspond to the (putative) assignment of the sequence of two peptides, connected by a crosslink at defined positions within the peptide sequences. This has implications on error handling, as will be discussed below. Therefore, the term crosslink-spectrum match might be more suitable. CSMs/XSMs may subsequently be collapsed into higher level contact information: peptide pairs, residue pairs or protein pairs. It should be noted that Crosslinking MS also has to address the protein inference problem of proteomics (Nesvizhskii and Aebersold, 2005; Rappsilber and Mann, 2002) due to the existence of multiple proteins with overlapping sequences. In addition, as multiple copies of the same protein are present in a sample, crosslinks of a protein to itself may be intramolecular or intermolecular. Often one cannot distinguish these self-links without dedicated experimental design or additional considerations (Lima et al., 2018; Taverner et al., 2002).

Identifications at all levels are associated with some error rate. This concerns the identity of the peptides (characterised by the false discovery rate, FDR) and the localisation of the crosslinking sites (characterised by the false localisation rate, FLR). FDRs can, in principle, be determined by so called target/decoy search strategies whereby the MS/MS spectra are searched against the target sequence database and a database of non-natural decoy sequences that are typically obtained by reversing or shuffling the sequences contained in the target database (Fischer and Rappsilber, 2017, 2018; Maiolica et al., 2007; Walzthoeni et al., 2012). The frequency of matches to the decoy database is then assumed to be equivalent to the frequency of random hits to the target database, this correlation is used to calculate the FDR. There are several caveats to consider for FDR control in crosslinking: First, the mere fact that a combination of two peptides is identified raises the chance for error compared to single peptide chain identifications. It suffices if one of the two peptides is false for the whole to be false (Trnka et al., 2014). Also, false positives increase proportionally when moving from CSM/XSM to peptide pair to protein pair level (Fischer and Rappsilber, 2017). This error propagation is a result of the typically observed redundancy of true positive hits (multiple CSMs/XSMs per peptide pair, multiple peptide pairs per protein pair) while random false positives by definition are less redundant. Therefore, FDRs need to be controlled at multiple levels, in the same way as for conventional proteomics experiments when moving from CSMs/XSMs to identified peptides and identified proteins. Second, the search space of self and heteromeric crosslinks are of different size. Therefore, the error of self and heteromeric links must be considered separately (Lenz et al., 2020; Walzthoeni et al., 2012). Third, for samples of limited complexity, when using only a small sequence database, and dependent on the crosslinking chemistry and data analysis strategy, there may be an insufficient number of decoy hits for an accurate determination of FDRs. Most FDR strategies try to model the tail of the false positive score distributions, but in a sparse data set this boundary is strongly affected by the selection of search parameters such as database composition (inclusion/exclusion of contaminant proteins) or the defined specificity of the crosslinking reagent.

Although the field has already seen substantial progress in FDR control, many commonly used software tools do not yet support FDR control at all levels, and there is no consensus yet for how FDRs should be collected for all experimental designs currently applied in crosslinking (Beveridge et al., 2020; Keller et al., 2019a; Yugandhar et al., 2020). Especially when dealing with error rates at the crosslinking site level, the FLR (related to the confidence of correctly assigning the crosslinked residues within the two peptide sequences) also needs to be taken into account. This is an even more challenging problem because precise localisation of the crosslink sites requires the observation of not just any fragments of the peptides but those that allow excluding alternative sites. Often, the linkage site is assigned based on the known or assumed reactivity of the crosslinker, an approach that fails at least when using photo-crosslinking (Schneider et al., 2018). Note that crosslinks reveal proximity and it remains to be seen what precision is required by modelling. Here, Crosslinking MS may differ from PTM mapping, for example, where the exact PTM site can be critical for mechanistic studies. Additional challenges arise when chromatographic co-elution of the same peptide pair but linked at different sites occurs.

Eventually, the outcome of crosslinking experiments is made publicly available through different channels, mostly journals and data repositories. Research articles provide experimental details and mostly qualitative but increasingly also quantitative crosslinking results (Chen and Rappsilber, 2018), in widely varying degrees of detail. Crosslinking identifications are typically reported in a tabular format in the main manuscript or (more commonly) integrated into the online supporting information section as standalone tables or formatted in a joint file with other supplementary data, for example in PDF format. Although this solution may at least fulfil minimum expectations of data transparency, such a deposition complicates data reuse and reanalysis especially as the formatting lacks a common standard that specifies what essential data should be included in such a table. Proteomics data repositories already offer some support for crosslinking data sets; for example, a project can be designated as a crosslinking study in the ProteomeXchange partner repository PRIDE and all data necessary for a "complete" submission can be provided. However, the submission is labelled as "partial" and therefore does give the false impression of being not entirely adhering to open data sharing principles and cannot be cited through a Digital Object Identifier, which is increasingly becoming part of open data policies (Gierasch et al., 2020). The PXC partner repository jPOST accepts such complete submissions as "complete", fortunately. Version 1.2 of the open mzIdentML standard (Vizcaíno et al., 2017) developed by the Proteomics Standard Initiative of the Human Proteome Organization and the official PSI validator offer support for some crosslinking results (Montecchi-Palazzi et al., 2009), but not all types of workflow are supported, for example the increasingly popular cleavable crosslinking reagents are not completely covered. The less complex mzTab format would be an alternative output format.

Apart from the MS-centric data deposition, alternative locations for data integration would be resources such as protein sequence databases (e.g. UniProt (UniProt Consortium, 2019)) or protein interaction databases such as IntAct (Orchard et al., 2014), STRING (Szklarczyk et al., 2019), BioGRID (Oughtred et al., 2019) or ComplexPortal (Meldal et al., 2019). In fact, IntAct is already including published crosslinking data on protein-protein interactions, even though the Crosslinking MS field has not established appropriate quality control mechanisms.

Some crosslinking data are also available through individual lab efforts such as XLinkDB (Keller et al., 2019b). In the specific context of the use of crosslinking data for integrative/hybrid modelling, the data dictionary (Vallat et al., 2018) used by PDB-Dev already offers support for crosslinking site centric distance restraints. However, at the moment there is no interoperability between these different resources that would seamlessly connect all these different repositories and databases. The following section will explain why this would be highly valuable to different audiences.

**4. Requirements for maximal impact of Crosslinking MS**

To maximise the use of Crosslinking MS its data should be made available adhering to the principles of FAIR (Findable, Accessible, Interoperable, Reusable) (Wilkinson et al., 2016). In addition, all of the experimental steps should be transparent to others by providing a sufficient amount of information, defined jointly by the community, and by providing this information in a suitable format in articles and in data repositories. This has been done for multiple other proteomics data types (http://www.psidev.info/miape) (Taylor et al., 2007). However, different communities that stand to benefit require different types of data and differ in the level of detail required for reuse. This needs to be considered before planning a course of action.

Peers (wet- and dry-lab scientists working in the crosslinking area) may use data to learn about new developments in the field, to assess the validity of published work and to reanalyse existing data sets, for example in the context of software development. For these purposes, detailed information and access to many different files are required. This includes raw/primary MS data and peak lists used for the initial search together with the search configuration and database. It also includes "technical" metadata (including details related to the original search such as software (version) and search parameters) and details about the instrumentation for which already a first example of a reporting template exists (Iacobucci et al., 2019). Finally, one also requires identifications at different levels (CSMs/XSMs, peptide pairs, residue/site pairs, protein pairs) including decoys, details about the FDR control (what approach was used, at which levels was FDR control applied, although this should ideally be standardised), and "biological" metadata (related to the nature of the sample and the experimental design, e.g. replicates or perturbations and sample treatment) together with the link to a publication if applicable.

Structural, computational or systems biologists are more likely to not work with the raw MS data themselves, but they will be rather interested in using the outcome of crosslinking experiments for modelling protein conformations, protein complexes or cellular networks. As a consequence, these communities may primarily be interested in the identifications at residue level or protein-protein interaction level with associated measures of confidence (FDR) and possibly abundance. The chemistry of the crosslinking reagent should be well defined regarding reactive sites and spacer length to define appropriate boundaries for crosslink restraints. A stable link to the primary data is required, for example in a proteomics repository, and the data needs to be provided in standardised form (also a wwPDB Integrative/Hybrid Methods Task Force recommendation (Berman et al., 2019; Sali et al., 2015)). The data should be findable and experimental details documented, i.e. some basic technical and biological metadata need to be associated with the data together with a link to a more detailed description, ideally a publication.

Finally, molecular and cell biologists and other researchers interested in protein interactions in general might be interested in crosslinking data because they represent binary interactions between proteins and/or specific residues in proteins. For these communities, the biggest value will come from access through an intuitive interface, to identifications at residue level or protein-protein interaction level with associated measures of confidence (FDR). This might be best achieved by the integration of such data into resources (databases) that they normally use, such as IntAct, STRING, or UniProt. A useful point of reference would be the HUPO PSI-MI standard which records molecular interactions without including the supporting MS data. These access points may either need to expand their data visualisation to include topological information or an additional interface may be needed that provides intuitive access also to residue-level information, akin to what is offered by tools such as xVis, xiNET and xiVIEW (Combe et al., 2015; Graham et al.; Grimm et al., 2015) or in field databases such as ProXL (Riffle et al., 2016).

In summary, different user bases require a different scope and granularity of the information that is obtained from crosslinking experiments. In any case, the ideal scenario would be a transparent and seamless flow of information to and from all resources connected to crosslinking in standardised formats, raising the question which parts of the workflow can and should be standardised.

**5. Recommendations as to where the Crosslinking MS field requires standardisation**

We feel that the Crosslinking MS field will benefit the most from field-developed standards in four specific areas of Crosslinking MS analyses and reporting, leading to the eleven tasks summarised in **Table 1** and presented in detail below.

| | Table 1. Recommendations (single sentence summaries of the field's to-do list) |
|---|---|
| 1 | Define best practices in experimental design for different applications of Crosslinking MS. |
| 2 | Find consensus on procedures to reliably assess error rates for all workflow types and at different levels (site pair to protein pair). |
| 3 | Ensure support by and complete integration with proteomics data repositories such as proteomeXchange. |
| 4 | Develop consistent terminology and common vocabularies for metadata annotation of Crosslinking MS data sets. |
| 5 | Provide enhanced support for data sharing with community-agreed file formats such as mzIdentML or mzTab. |
| 6 | Define minimal requirements for reporting Crosslinking MS data in peer-reviewed publications. |
| 7 | Facilitate access to modellers by providing results in formats suitable for structure and model repositories such as wwPDB/PDB-Dev. |
| 8 | Develop parsers for data integration in interaction databases and develop easily accessible visualisation tools. |
| 9 | Ensure flexibility for new developments in the field, not all steps need to be standardised as workflows evolve. |
| 10 | Organise benchmarking studies for objective comparisons of key experimental and computational steps. |
| 11 | Establish minimum reporting standards for reporting new or improved reagents and software tools. |

## 5.1 Workflows / Experiment

**Recommendation 1: Best practice in experimental design**

While there is a large diversity of analytical tools and concepts being utilised in Crosslinking MS they are all based on the same principle of preserving structural information by introducing artificial covalent bonds in and between biomolecules that would otherwise be lost during the mass spectrometric analysis. Therefore guidelines should be developed ensuring that the resulting data can conclusively be interpreted. This should address fundamental aspects of experimental design such as the number and type of replicates and whether different recommendations are required for different types of experiments, e.g. the analysis of highly purified, individual proteins or small protein complexes versus whole cell analysis and qualitative versus quantitative experiments. This may include control experiments to address oligomerization or sample integrity. An appropriate mechanism has to be set up that allows identifying where best practice guidelines are needed and then developing these guidelines while allowing their continuous adaptation to our progressing understanding of Crosslinking MS.

**Recommendation 2: Error assessment**

It is important to determine the error of Crosslinking MS data by a transparent and thoroughly tested method. There is currently a large number of methods for FDR control that are usually based on the target-decoy approach. Also comparisons are made to available high-resolution structures, which has its limitations as this is also experimental data and in addition describes a static representation of a protein or protein complex. It is of utmost importance that the field arrives at a consensus for procedures that return a reliable error assessment. It is also important that the limits of these procedures are mapped out. Future studies hopefully then employ this field-agreed error assessment method in its respective current form. This would be largely helped by swift integration into the main data analysis workflows by the respective developers. Changes of the procedure must be well documented and thoroughly tested before coming into effect.

## 5.2 Data sharing

**Recommendation 3: Public repositories**

All data of Crosslinking MS should be shared in an open and stable way in a public repository. This provides a stable dataset identifier such as an accession number and a defined data structure to cross-reference crosslinking datasets in other resources and requires standardised metadata (Recommendation 4) and standardised file formats (Recommendation 5). This is already the case for proteomics data and there is no need to develop a Crosslinking MS specific new repository, although Crosslinking MS specific adaptations of existing proteomics repositories are needed. The PRIDE repository, as one of the PXC members, has committed to being a partner in this endeavour. Basic functionality for "partial" submissions (raw MS data and some metadata) is already available, but should be considered the minimum baseline for data sharing. For "complete" submission, some extensions need to be made to better address the Crosslinking MS results. Another PXC member, the jPOST repository, has already accepted several Crosslinking MS projects with "complete submission", but will continuously need to work with the field to universally address the diverse data modalities of Crosslinking MS. Ultimately, the criteria for "complete" submission are Crosslinking MS specific and need to be defined by the field. These then must be implemented through basic and in the longer term possibly also more elaborate checks during submission, making "complete" submissions possible for Crosslinking MS data independent of criteria applied to the data of other fields.

The Crosslinking MS data that enters public repositories should receive a quality check at all levels, ideally automatically at the point of uploading. This pertains to the elemental integrity of the files, their adherence to the agreed standard formats which includes semantic validation and readability by parsers that increase the data availability and reach, and data quality metrics such as a measure of confidence. For this, appropriate software will need to be developed and maintained in a field effort, in collaboration with data repositories. Results of Crosslinking MS that then enter other repositories should do so together with a measure of confidence (see Recommendation 8).

**Recommendation 4: Metadata**

All information needed to reproduce Crosslinking MS results must be provided in full. Duplications in locations where this takes place should be minimised, though. A minimal set of critical information that is required for a basic understanding of the Crosslinking MS results should be provided as part of data submission. A standardised description of a crosslinking experiment requires the definition of common, controlled vocabularies. XLMOD (Mayer, 2020) (https://raw.githubusercontent.com/HUPO-PSI/mzIdentML/master/cv/XLMOD.obo) as an effort coordinated via the HUPO PSI regarding controlled vocabularies in crosslinking covers "cross-linking reagents, cross-linker related post-translational modifications, and derivatisation reagents for GC-MS and LC-MS". Other terms for a standardised, minimal description of a Crosslinking MS experiment will need to be defined in additional efforts. Full experimental details should be provided in the experimental section of publications. A good starting point of what should be included here are the recommendations emerging from a first community study (Iacobucci et al., 2019). This is currently taking a tabular form as is also practised in other fields (Henderson et al., 2012; Masson et al., 2019; Montelione et al., 2013; Read et al., 2011; Trewhella et al., 2017). Metadata about instrumentation and data acquisition parameters would ideally be parsed from the raw data before submission and written automatically into the submitted file. Likewise, search parameters should automatically be documented in a form so that users can effortlessly pass them on to the data repository as part of their submission. It would be desirable to minimise the number of separate files by combining all relevant experimental information possibly together with the results into a single file.

**Recommendation 5: Community-agreed file formats**

All data that is shared should be shared in an open and community-agreed format that is extensible to support the evolving needs of the community. This has been successfully performed for peak list formats such as mzML. A standard result file format should be developed and include a complete list of target and decoy identifications as potential true and known noise distributions. Integrative/hybrid modelling is tolerant to substantial error rates and including the known noise (decoy matches) allows the modelling field to build ways how to deal with noise in crosslinking data into their procedures (Berman et al., 2019; Rout and Sali, 2019) although the decoy matches form only an initial noise model that is converted into an actual noise model by FDR estimation procedures. This format might be based on already existing standard formats such as mzIdentML or mzTab, which are supported by the HUPO-PSI as the initiative in the proteomics field on procedures for standardisation. mzIdentML 1.2 would be a starting point for further efforts since it already supports some albeit not all types of crosslinking data (Vizcaíno et al., 2017). Crosslinking data is currently not supported in mzTab, but also mzTab could be extended to accommodate this data. In addition to output formats one should also keep an eye on input files and their standardisation, which includes mzML for peak lists (Martens et al., 2011) and PEFF for sequence databases (Binz et al., 2019). In any case, part of integrating the needs of the crosslinking community into general proteomics standards will be developing parser libraries, readers and writers. Crosslinking MS search software should be adapted to write results and search parameters in standards compliant form to allow direct sharing of data, metadata and results. As crosslinking is evolving as a method this will also lead to evolving standards and result in a continuous need to update software tools. We acknowledge that the complexity of the data makes the whole process of changing existing software tools and also maintaining them challenging.

**Recommendation 6: Publication guidelines**

Publication guidelines should be developed for what constitutes a sufficiently detailed description of experimental design (Recommendation 1), sample and data processing (Recommendations 2 and 4), and presentation of results (Recommendations 3 and 5).

**5.3 Knowledge transfer**

**Recommendation 7: Access to Crosslinking MS results for modelling**

Efforts should be undertaken to maximise access of other researchers and communities to the link and interaction data obtained by Crosslinking MS. In fact, wwPDB/PDB-Dev has reached out in the name of the structural biology and modelling field to the Crosslinking MS field with the specific requirement of having access to Crosslinking MS data held in a public repository in standardised form and with quality descriptors. These requirements are going to be met by Recommendations 2 (FDR), 3 (data repository) and 5 (file formats). Once Crosslinking MS formats have been established, parsers can and will be written to stably link the data into the workflows of the modelling community and the public model repository, PDBdev.

**Recommendation 8: Accessibility of Crosslinking MS data to experimentalists**

Providing biological researchers with efficient access to Crosslinking MS results requires those data to be integrated in existing resources containing protein-protein interaction information such as IntAct, UniProt and STRING. This requires first of all Crosslinking MS data to make it into these repositories in an automated and quality-controlled way (linking to Recommendation 2 and 3). This includes the writing of parsers that convert Crosslinking MS results into formats for molecular interactions and protein complexes. This also mandates the further development of crosslinking data visualisation tools and their integration with public databases. These visualisation tools can be broadly categorised by their purpose: Investigating spectral data,protein structure, or protein interaction networks. This does not necessarily comprise a definitive listing as these tools and their integration with each other and other tools is under active development. Even within the first category, consisting of spectral interpretations, the often long lists of crosslinked proteins and crosslinked amino acid residues returned by Crosslinking MS are not intuitively understandable to humans. The ability to display residue-resolution information provided by Crosslinking MS has been shown to provide a more suitable visual data interaction platform ((Combe et al., 2015; Graham et al.; Hoopmann et al., 2015, 2016; Keller et al., 2019b; Kolbowski et al., 2018; Riffle et al., 2016), among a subset of examples). Tools increase in value through integration, for example node-link diagrams that classically display protein interaction data (Combe et al., 2015) can be supplemented by a display of the residue-level spectral interpretations (Graham et al.; Riffle et al., 2016). Visualisation tools should be (further) developed to allow a seamless interrogation of Crosslinking MS data from a wide angle of perspectives with a focus on understanding the data and developing testable hypotheses from it. This requires linking of the visualisation to the public repository of Crosslinking MS data on one side and public repositories of protein function and interaction data on the other side. Visualisation tools should have low entry barriers such as being browser-based (Combe et al., 2015; Deutsch et al., 2015; Graham et al.; Kolbowski et al., 2018; Riffle et al., 2016; Trnka et al., 2014) or easy to install (Kosinski et al., 2015), and be open source and grants funded to ensure transparent development and access by the widest possible number of researchers.

## 5.4 Future development of Crosslinking MS

**Recommendation 9: Crosslinking MS comes in many flavours**

Crosslinking MS is currently seeing the rapid prototyping of novel workflows. These workflows implement different ideas around the same basic concept but use in part very different analytical tools. Given the diversity of approaches that exist for proteomics, it is unclear if ever a unified workflow will arise for Crosslinking MS (Leitner et al., 2016; O'Reilly and Rappsilber, 2018; Sinz, 2018; Steigenberger et al., 2020; Yu and Huang, 2018). Therefore, many fundamental elements of the workflow *should not* be subject to strict standardisation at this point. This specifically includes (a) crosslinking reagents, as many types of crosslinking reagents and chemistries exist and new ones are introduced on a regular basis; (b) instrumentation, as many types of mass spectrometers with diverse features (e.g. fragmentation techniques, real-time decision making processes) exist and certainly the technology will continue to evolve; and (c) data analysis software, as many types of crosslinking analysis software exist and, again, there is a continuous development of entirely new software tools or new versions of existing tools.

In addition, there are many different ways of combining chemistry, MS and bioinformatics, although dependencies can and do exist (e.g., a software will only work with crosslinking reagents of a certain design, or it may only accept certain types of MS data). Although some workflows may be more suitable than others for a given application, crosslinking can be applied in many different contexts. There is a large diversity of strategies being explored and will be explored for the foreseeable future. Clearly, we believe that a larger number of workflow designs will be fit for purpose. Note that diversity also exists in other fields, for example many different software tools are used successfully for protein identification in proteomics.

**Recommendation 10: Community benchmark exercise**

The first community-wide, comparative Crosslinking MS study published in 2019 (Iacobucci et al., 2019) highlighted the desire of the field for transparent assessments of the many different workflows through organised challenges. This should be continued and expanded to include all application areas of Crosslinking MS from single proteins over multi-protein complexes to highly complex mixtures of proteins. These challenges should provide both experimentalists and computational scientists the opportunity to showcase and benchmark their tools and demonstrate the progress of the field or/and highlight remaining challenges in the respective areas.

**Recommendation 11: Minimal standards for the reporting of new tools**

New crosslinking reagents and new search software (versions) are frequently reported. While these reports typically contain proof-of-principle evidence they often lack data that allows assessing in full the merits of these new or changed tools. This limits not only the uptake of these tools but also makes it more difficult for others to plan their experiments in light of the many choices that are on offer. We should therefore develop guidelines and possibly benchmark challenges that provide the field with a general comparability of tools and ideally a quantitative assessment of progress. The above-mentioned organised challenges are one approach to this, albeit infrequent and should be supplemented by rolling or/and fixed challenges. Rolling challenges are known for example in protein structure modelling: Continuous Automated Model Evaluation (CAMEO) (Haas et al., 2018). Here an available yet confidential structure is used as ground truth against which submitted models are assessed. Especially when it comes to protein-protein interactions such a ground truth typically does not exist for Crosslinking MS and therefore alternative approaches for evaluation will need to be developed.

## 6. Implementation

Crowd-sourcing community standards is known to be a very time-consuming process. To streamline this process, we suggest that initially a small group gathered from the authors of this paper and any other interested party (please contact AL or JR) proposes such standards. Initial discussions may be performed through online discussion groups, where different opinions towards best practices can be shared and specific challenges discussed. Once a consensus emerges, recommendations would be presented at a community meeting such as the annual Symposium on Structural Proteomics and reported in a publication. This process is to some extent reminiscent of the procedures established by the HUPO-PSI. Many of the different aspects of standardisation will require significant funding, for example, to develop new and adapt existing software, and continuous funding support to ensure standards and their implementation tools to evolve with the changing needs of this rapidly developing field.

**Acknowledgements**

We thank the participants of the meetings mentioned throughout the text for their many different contributions. While we made an effort to include all labs that may feel belonging to the field of Crosslinking MS we apologise to all those that we missed. Developing standards in Crosslinking MS is to be viewed as an inclusive activity, so please feel warmly invited to join our efforts.

This work was supported by the Wellcome Trust through a Senior Research Fellowship to JR (103139) and the Deutsche Forschungsgemeinschaft (DFG, German Research Foundation, project no. 329673113). Funded by the Deutsche Forschungsgemeinschaft (DFG, German Research Foundation) under Germany´s Excellence Strategy – EXC 2008/1 – 390540038. The Wellcome Centre for Cell Biology is supported by core funding from the Wellcome Trust (203149).

**Author contributions**

Writing – Original draft, A.L. and J.R., Writing – Review and Editing, all authors.

**Declaration of Interests**

The authors declare no competing interests.

**References**

Berman, H.M., Burley, S.K., Chiu, W., Sali, A., Adzhubei, A., Bourne, P.E., Bryant, S.H., Dunbrack, R.L., Jr, Fidelis, K., Frank, J., et al. (2006). Outcome of a workshop on archiving structural models of biological macromolecules. Structure *14*, 1211–1217.

Berman, H.M., Adams, P.D., Bonvin, A.A., Burley, S.K., Carragher, B., Chiu, W., DiMaio, F., Ferrin, T.E., Gabanyi, M.J., Goddard, T.D., et al. (2019). Federating Structural Models and Data: Outcomes from A Workshop on Archiving Integrative Structures. Structure *27*, 1745–1759.

Beveridge, R., Stadlmann, J., Penninger, J.M., and Mechtler, K. (2020). A synthetic peptide library for benchmarking crosslinking-mass spectrometry search engines for proteins and protein complexes. Nat. Commun. *11*, 742.

Binz, P.-A., Shofstahl, J., Vizcaíno, J.A., Barsnes, H., Chalkley, R.J., Menschaert, G., Alpi, E., Clauser, K., Eng, J.K., Lane, L., et al. (2019). Proteomics Standards Initiative Extended FASTA Format. J. Proteome Res. *18*, 2686–2692.

Burley, S.K., Kurisu, G., Markley, J.L., Nakamura, H., Velankar, S., Berman, H.M., Sali, A., Schwede, T., and Trewhella, J. (2017). PDB-Dev: a Prototype System for Depositing Integrative/Hybrid Structural Models. Structure *25*, 1317–1318.

Chen, Z.A., and Rappsilber, J. (2018). Protein Dynamics in Solution by Quantitative Crosslinking/Mass Spectrometry. Trends Biochem. Sci. *43*, 908–920.

Combe, C.W., Fischer, L., and Rappsilber, J. (2015). xiNET: cross-link network maps with residue resolution. Mol. Cell. Proteomics *14*, 1137–1147.

Deutsch, E.W., Mendoza, L., Shteynberg, D., Slagel, J., Sun, Z., and Moritz, R.L. (2015). Trans-Proteomic Pipeline, a standardized data processing pipeline for large-scale reproducible proteomics informatics. Proteomics Clin. Appl. *9*, 745–754.

Deutsch, E.W., Orchard, S., Binz, P.-A., Bittremieux, W., Eisenacher, M., Hermjakob, H., Kawano, S., Lam, H., Mayer, G., Menschaert, G., et al. (2017). Proteomics Standards Initiative: Fifteen Years of Progress and Future Work. J. Proteome Res. *16*, 4288–4298.

Deutsch, E.W., Bandeira, N., Sharma, V., Perez-Riverol, Y., Carver, J.J., Kundu, D.J., García-Seisdedos, D., Jarnuczak, A.F., Hewapathirana, S., Pullman, B.S., et al. (2020). The ProteomeXchange consortium in 2020: enabling "big data" approaches in proteomics. Nucleic Acids Res. *48*, D1145–D1152.

Fischer, L., and Rappsilber, J. (2017). Quirks of Error Estimation in Cross-Linking/Mass Spectrometry. Anal. Chem. *89*, 3829–3833.

Fischer, L., and Rappsilber, J. (2018). False discovery rate estimation and heterobifunctional cross-linkers. PLoS One *13*, e0196672.

Gierasch, L.M., Davidson, N.O., Rye, K.-A., and Burlingame, A.L. (2020). The Data Must Be Accessible to All. Mol. Cell. Proteomics *19*, 569–570.

Graham, M.J., Combe, C., Kolbowski, L., and Rappsilber, J. xiView: A common platform for the downstream analysis of Crosslinking Mass Spectrometry data. bioRxiv, DOI: 10.1101/561829.

Grimm, M., Zimniak, T., Kahraman, A., and Herzog, F. (2015). xVis: a web server for the schematic visualization and interpretation of crosslink-derived spatial restraints. Nucleic Acids Res. *43*, W362–W369.

Haas, J., Barbato, A., Behringer, D., Studer, G., Roth, S., Bertoni, M., Mostaguir, K., Gumienny, R., and Schwede, T. (2018). Continuous Automated Model EvaluatiOn (CAMEO) complementing the critical assessment of structure prediction in CASP12. Proteins *86 Suppl 1*, 387–398.

Henderson, R., Sali, A., Baker, M.L., Carragher, B., Devkota, B., Downing, K.H., Egelman, E.H., Feng, Z., Frank, J., Grigorieff, N., et al. (2012). Outcome of the first electron microscopy validation task force meeting. Structure *20*, 205–214.

Hoopmann, M.R., Zelter, A., Johnson, R.S., Riffle, M., MacCoss, M.J., Davis, T.N., and Moritz, R.L. (2015). Kojak: efficient analysis of chemically cross-linked protein complexes. J. Proteome Res. *14*, 2190–2198.

Hoopmann, M.R., Mendoza, L., Deutsch, E.W., Shteynberg, D., and Moritz, R.L. (2016). An Open Data Format for Visualization and Analysis of Cross-Linked Mass Spectrometry Results. J. Am. Soc. Mass Spectrom. *27*, 1728–1734.

Iacobucci, C., Piotrowski, C., Aebersold, R., Amaral, B.C., Andrews, P., Bernfur, K., Borchers, C., Brodie, N.I., Bruce, J.E., Cao, Y., et al. (2019). First Community-Wide, Comparative Cross-Linking Mass Spectrometry Study. Anal. Chem. *91*, 6953–6961.

Keller, A., Chavez, J.D., Felt, K.C., and Bruce, J.E. (2019a). Prediction of an Upper Limit for the Fraction of Interprotein Cross-Links in Large-Scale In Vivo Cross-Linking Studies. J. Proteome Res. *18*, 3077–3085.

Keller, A., Chavez, J.D., Eng, J.K., Thornton, Z., and Bruce, J.E. (2019b). Tools for 3D Interactome Visualization. J. Proteome Res. *18*, 753–758.

Kolbowski, L., Combe, C., and Rappsilber, J. (2018). xiSPEC: web-based visualization, analysis and sharing of proteomics data. Nucleic Acids Res. *46*, W473–W478.

Kosinski, J., von Appen, A., Ori, A., Karius, K., Müller, C.W., and Beck, M. (2015). Xlink Analyzer: software for analysis and visualization of cross-linking data in the context of three-dimensional structures. J. Struct. Biol. *189*, 177–183.

Lawson, C.L., Baker, M.L., Best, C., Bi, C., Dougherty, M., Feng, P., van Ginkel, G., Devkota, B., Lagerstedt, I., Ludtke, S.J., et al. (2011). EMDataBank.org: unified data resource for CryoEM. Nucleic Acids Res. *39*, D456–D464.

Leitner, A., Faini, M., Stengel, F., and Aebersold, R. (2016). Crosslinking and Mass Spectrometry: An Integrated Technology to Understand the Structure and Function of Molecular Machines. Trends Biochem. Sci. *41*, 20–32.

Lenz, S., Sinn, L.R., O'Reilly, F.J., Fischer, L., Wegner, F., and Rappsilber, J. (2020). Reliable identification of protein-protein interactions by crosslinking mass spectrometry. bioRxiv, DOI: 10.1101/2020.05.25.114256.

Lima, D.B., Melchior, J.T., Morris, J., Barbosa, V.C., Chamot-Rooke, J., Fioramonte, M., Souza, T.A.C.B., Fischer, J.S.G., Gozzo, F.C., Carvalho, P.C., et al. (2018). Characterization of homodimer interfaces with cross-linking mass spectrometry and isotopically labeled proteins. Nat. Protoc. *13*, 431–458.

Maiolica, A., Cittaro, D., Borsotti, D., Sennels, L., Ciferri, C., Tarricone, C., Musacchio, A., and Rappsilber, J. (2007). Structural analysis of multiprotein complexes by cross-linking, mass spectrometry, and database searching. Mol. Cell. Proteomics *6*, 2200–2211.

Martens, L., Chambers, M., Sturm, M., Kessner, D., Levander, F., Shofstahl, J., Tang, W.H., Römpp, A., Neumann, S., Pizarro, A.D., et al. (2011). mzML—a Community Standard for Mass Spectrometry Data. Mol. Cell. Proteomics *10*, R110.000133.

Masson, G.R., Burke, J.E., Ahn, N.G., Anand, G.S., Borchers, C., Brier, S., Bou-Assaf, G.M., Engen, J.R., Englander, S.W., Faber, J., et al. (2019). Recommendations for performing, interpreting and reporting hydrogen deuterium exchange mass spectrometry (HDX-MS) experiments. Nat. Methods *16*, 595–602.

Mayer, G. (2020). XLMOD: Cross-linking and chromatography derivatization reagents ontology. arXiv:2003.00329. Available at: https://arxiv.org/abs/2003.00329.

Meldal, B.H.M., Bye-A-Jee, H., Gajdoš, L., Hammerová, Z., Horácková, A., Melicher, F., Perfetto, L., Pokorný, D., Lopez, M.R., Türková, A., et al. (2019). Complex Portal 2018: extended content and enhanced visualization tools for macromolecular complexes. Nucleic Acids Res. *47*, D550–D558.

Montecchi-Palazzi, L., Kerrien, S., Reisinger, F., Aranda, B., Jones, A.R., Martens, L., and Hermjakob, H. (2009). The PSI semantic validator: a framework to check MIAPE compliance of proteomics data. Proteomics *9*, 5112–5119.

Montelione, G.T., Nilges, M., Bax, A., Güntert, P., Herrmann, T., Richardson, J.S., Schwieters, C.D., Vranken, W.F., Vuister, G.W., Wishart, D.S., et al. (2013). Recommendations of the wwPDB NMR Validation Task Force. Structure *21*, 1563–1570.

Moriya, Y., Kawano, S., Okuda, S., Watanabe, Y., Matsumoto, M., Takami, T., Kobayashi, D., Yamanouchi, Y., Araki, N., Yoshizawa, A.C., et al. (2019). The jPOST environment: an integrated proteomics data repository and database. Nucleic Acids Res. *47*, D1218–D1224.

Nesvizhskii, A.I., and Aebersold, R. (2005). Interpretation of shotgun proteomic data: the protein inference problem. Mol. Cell. Proteomics *4*, 1419–1440.

Orchard, S., Ammari, M., Aranda, B., Breuza, L., Briganti, L., Broackes-Carter, F., Campbell, N.H., Chavali, G., Chen, C., del-Toro, N., et al. (2014). The MIntAct project--IntAct as a common curation platform for 11 molecular interaction databases. Nucleic Acids Res. *42*, D358–D363.

O'Reilly, F.J., and Rappsilber, J. (2018). Cross-linking mass spectrometry: methods and applications in structural, molecular and systems biology. Nat. Struct. Mol. Biol. *25*, 1000–1008.

Oughtred, R., Stark, C., Breitkreutz, B.-J., Rust, J., Boucher, L., Chang, C., Kolas, N., O'Donnell, L., Leung, G., McAdam, R., et al. (2019). The BioGRID interaction database: 2019 update. Nucleic Acids Res. *47*, D529–D541.

Perez-Riverol, Y., Csordas, A., Bai, J., Bernal-Llinares, M., Hewapathirana, S., Kundu, D.J., Inuganti, A., Griss, J., Mayer, G., Eisenacher, M., et al. (2019). The PRIDE database and related tools and resources in 2019: improving support for quantification data. Nucleic Acids Res. *47*, D442–D450.

Rappsilber, J., and Mann, M. (2002). What does it mean to identify a protein in proteomics?

Trends Biochem. Sci. *27*, 74–78.

Read, R.J., Adams, P.D., Arendall, W.B., 3rd, Brunger, A.T., Emsley, P., Joosten, R.P., Kleywegt, G.J., Krissinel, E.B., Lütteke, T., Otwinowski, Z., et al. (2011). A new generation of crystallographic validation tools for the protein data bank. Structure *19*, 1395–1412.

Riffle, M., Jaschob, D., Zelter, A., and Davis, T.N. (2016). ProXL (Protein Cross-Linking Database): A Platform for Analysis, Visualization, and Sharing of Protein Cross-Linking Mass Spectrometry Data. J. Proteome Res. *15*, 2863–2870.

Rout, M.P., and Sali, A. (2019). Principles for Integrative Structural Biology Studies. Cell *177*, 1384–1403.

Sali, A., Berman, H.M., Schwede, T., Trewhella, J., Kleywegt, G., Burley, S.K., Markley, J., Nakamura, H., Adams, P., Bonvin, A.M.J.J., et al. (2015). Outcome of the First wwPDB Hybrid/Integrative Methods Task Force Workshop. Structure *23*, 1156–1167.

Schneider, M., Belsom, A., and Rappsilber, J. (2018). Protein Tertiary Structure by Crosslinking/Mass Spectrometry. Trends Biochem. Sci. *43*, 157–169.

Schwede, T., Sali, A., Honig, B., Levitt, M., Berman, H.M., Jones, D., Brenner, S.E., Burley, S.K., Das, R., Dokholyan, N.V., et al. (2009). Outcome of a workshop on applications of protein models in biomedical research. Structure *17*, 151–159.

Sinz, A. (2018). Cross-Linking/Mass Spectrometry for Studying Protein Structures and Protein-Protein Interactions: Where Are We Now and Where Should We Go from Here? Angew. Chem. Int. Ed Engl. *57*, 6390–6396.

Steigenberger, B., Albanese, P., Heck, A.J.R., and Scheltema, R.A. (2020). To Cleave or Not To Cleave in XL-MS? J. Am. Soc. Mass Spectrom. *31*, 196–206.

Szklarczyk, D., Gable, A.L., Lyon, D., Junge, A., Wyder, S., Huerta-Cepas, J., Simonovic, M., Doncheva, N.T., Morris, J.H., Bork, P., et al. (2019). STRING v11: protein-protein association networks with increased coverage, supporting functional discovery in genome-wide experimental datasets. Nucleic Acids Res. *47*, D607–D613.

Taverner, T., Hall, N.E., O'Hair, R.A.J., and Simpson, R.J. (2002). Characterization of an antagonist interleukin-6 dimer by stable isotope labeling, cross-linking, and mass spectrometry. J. Biol. Chem. *277*, 46487–46492.

Taylor, C.F., Paton, N.W., Lilley, K.S., Binz, P.-A., Julian, R.K., Jr, Jones, A.R., Zhu, W.,

Apweiler, R., Aebersold, R., Deutsch, E.W., et al. (2007). The minimum information about a proteomics experiment (MIAPE). Nat. Biotechnol. *25*, 887–893.

Trewhella, J., Hendrickson, W.A., Kleywegt, G.J., Sali, A., Sato, M., Schwede, T., Svergun, D.I., Tainer, J.A., Westbrook, J., and Berman, H.M. (2013). Report of the wwPDB Small-Angle Scattering Task Force: data requirements for biomolecular modeling and the PDB. Structure *21*, 875–881.

Trewhella, J., Duff, A.P., Durand, D., Gabel, F., Guss, J.M., Hendrickson, W.A., Hura, G.L., Jacques, D.A., Kirby, N.M., Kwan, A.H., et al. (2017). 2017 publication guidelines for structural modelling of small-angle scattering data from biomolecules in solution: an update. Acta Crystallogr D Struct Biol *73*, 710–728.

Trnka, M.J., Baker, P.R., Robinson, P.J.J., Burlingame, A.L., and Chalkley, R.J. (2014). Matching cross-linked peptide spectra: only as good as the worse identification. Mol. Cell. Proteomics *13*, 420–434.

UniProt Consortium (2019). UniProt: a worldwide hub of protein knowledge. Nucleic Acids Res. *47*, D506–D515.

Vallat, B., Webb, B., Westbrook, J.D., Sali, A., and Berman, H.M. (2018). Development of a Prototype System for Archiving Integrative/Hybrid Structure Models of Biological Macromolecules. Structure *26*, 894–904.e2.

Vizcaíno, J.A., Mayer, G., Perkins, S., Barsnes, H., Vaudel, M., Perez-Riverol, Y., Ternent, T., Uszkoreit, J., Eisenacher, M., Fischer, L., et al. (2017). The mzIdentML Data Standard Version 1.2, Supporting Advances in Proteome Informatics. Mol. Cell. Proteomics *16*, 1275–1285.

Walzthoeni, T., Claassen, M., Leitner, A., Herzog, F., Bohn, S., Förster, F., Beck, M., and Aebersold, R. (2012). False discovery rate estimation for cross-linked peptides identified by mass spectrometry. Nat. Methods *9*, 901–903.

Wilkinson, M.D., Dumontier, M., Aalbersberg, I.J.J., Appleton, G., Axton, M., Baak, A., Blomberg, N., Boiten, J.-W., da Silva Santos, L.B., Bourne, P.E., et al. (2016). The FAIR Guiding Principles for scientific data management and stewardship. Sci Data *3*, 160018.

Yu, C., and Huang, L. (2018). Cross-Linking Mass Spectrometry: An Emerging Technology for Interactomics and Structural Biology. Anal. Chem. *90*, 144–165.

Yugandhar, K., Wang, T.-Y., Leung, A.K.-Y., Lanz, M.C., Motorykin, I., Liang, J., Shayhidin,

E.E., Smolka, M.B., Zhang, S., and Yu, H. (2020). MaXLinker: Proteome-wide Cross-link Identifications with High Specificity and Sensitivity. Mol. Cell. Proteomics *19*, 554–568.

Zahn, V.H., and Meienhofer, J. (1958). Reaktionen von 1,5-difluor-2,4-dinitrobenzol mit insulin 2. Mitt. Versuche mit insulin. Makromol. Chem. *26*, 153–166.

**Figure legends**

**Figure 1. General Crosslinking MS workflow. a**, Crosslinkers comprise various chemistries and spacer lengths. Depending on the experimental workflow used, the crosslinker spacer may be cleavable in the mass spectrometer (C) or isotope labeled (star), or have moieties that can be biochemically enriched (E) and chemically released (R). **b**, Concentrations and reaction times must be empirically tested for each application to achieve optimal amounts of crosslinking. **c**, Proteins can be digested in solution or in gel to produce a mixture of crosslinked and linear peptides. Also "dead-end" products, where the crosslinker has hydrolysed on one end, "loop-links", where the crosslink ends up in a single peptide, and higher order products, comprising more than two peptides and/or more than one crosslinker moiety, can form. **d**, After digestion, crosslinked peptides are often enriched through chromatographic methods, such as size-exclusion chromatography, strong-cation exchange chromatography or affinity chromatography. **e**, MS/MS acquisition pipelines have been designed to increase the likelihood of selecting crosslinked peptide precursors for fragmentation. **f**, Various search software solutions have been developed to identify the two linked peptides from the spectra. **g**, Through methods that determine the false-discovery rate (FDR), the list of matches is cut to the desired confidence. **h**, The links are visualised and/or **i**, used as part of integrative modelling. **j**, The data are deposited in public repositories. Figure in parts adapted from (O'Reilly and Rappsilber, 2018)

**Figure 1.**

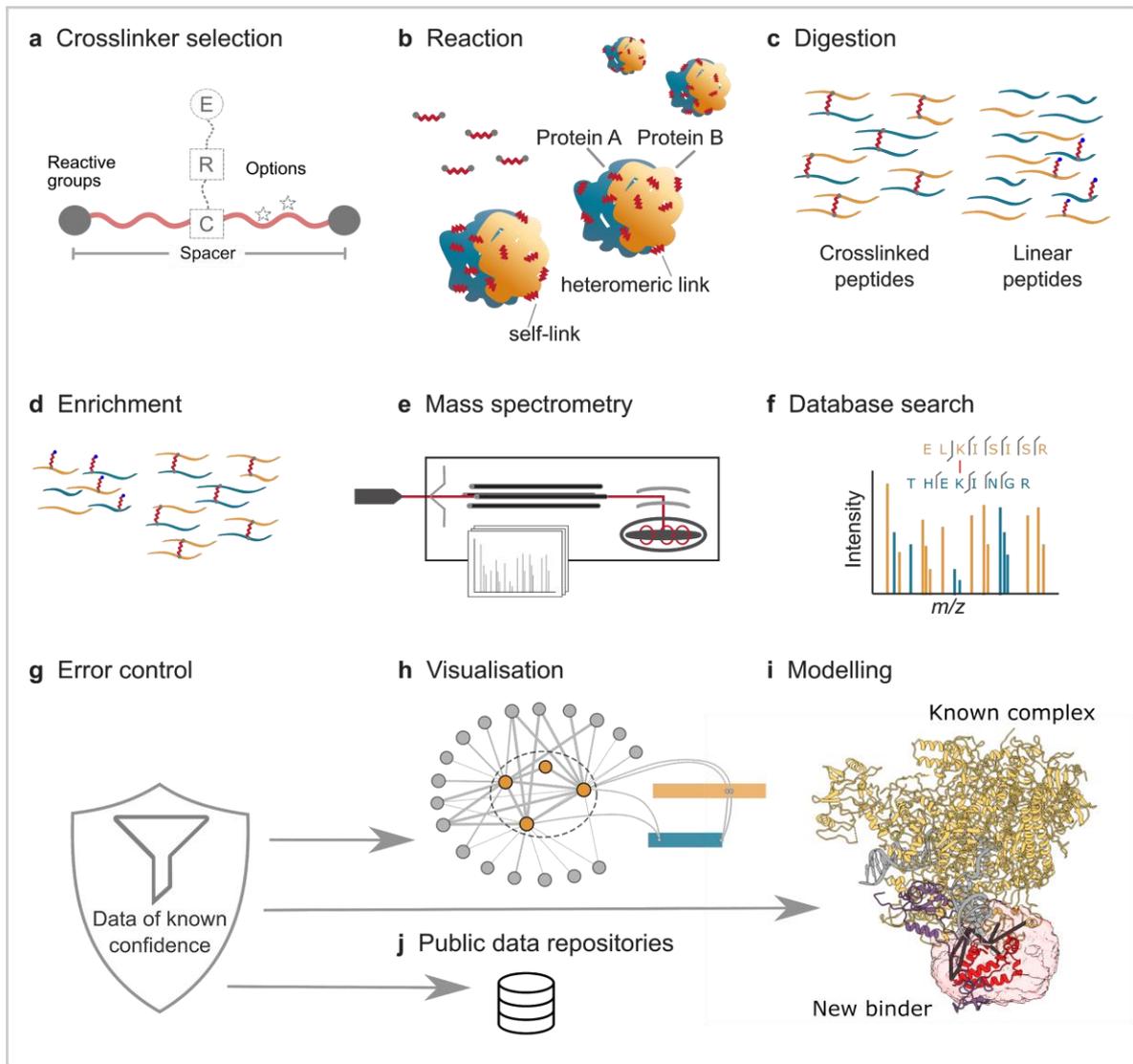